\documentclass{iopart}

\usepackage[T1]{fontenc}
\usepackage{ae,aecompl}

\usepackage[latin9]{inputenc}
\usepackage{graphicx}
\usepackage{amssymb}

\begin{document}

\title{On the physical properties of memristive, memcapacitive, and meminductive systems}

\author{Massimiliano Di Ventra}

\address{Department of Physics, University of California, San Diego, California 92093-0319, USA}
\ead{diventra@physics.ucsd.edu}

\author{Yuriy V. Pershin}
\address{Department of Physics and Astronomy and University of South Carolina Nanocenter, University of South Carolina, Columbia, South Carolina 29208, USA}
\ead{pershin@physics.sc.edu}

\begin{abstract}
We discuss the {\it physical} properties of realistic memristive, memcapacitive and meminductive systems.
In particular, by employing the well-known theory of response functions and microscopic derivations, we show that resistors, capacitors and inductors with memory emerge naturally in the response of systems - especially those of nanoscale dimensions - subjected to external perturbations.
As a consequence, since memristances, memcapacitances, and meminductances are simply response functions, they are not  necessarily finite. This means that, unlike what has always been argued in some literature,
{\it diverging} and {\it non-crossing} input-output curves of all these memory elements are physically possible in both quantum
and classical regimes. For similar reasons, it is not surprising to find memcapacitances and meminductances that
acquire {\it negative} values at certain times during dynamics, while the passivity criterion of memristive systems
imposes always a non-negative value on the resistance at any given time. We finally show that {\it ideal} memristors, namely those whose state depends only on the charge that flows through them (or on the history of the voltage) are subject to very strict physical conditions and are unable to protect their memory state against the unavoidable
fluctuations, and therefore are susceptible to a {\it stochastic catastrophe}. Similar considerations apply to ideal memcapacitors and meminductors.
\end{abstract}

\maketitle

\section{Introduction}

There is currently much interest about what we now call memristors, memcapacitors and meminductors, namely
 resistors, capacitors and inductors with memory \cite{chua71a,chua76a,diventra09a}. This interest is not just due to the arguably catchy name these elements have. A real and obvious advantage ensues if we have two-terminal devices that store information without a power source: they may replace -- at least in what concerns specific tasks regarding the manipulation and
 storing of information -- the transistor, a hallmark of our present-day microelectronics. However, the interest
 does not end here. These elements, when combined in complex circuits, can perform logic \cite{borghetti10a} and {\it non-traditional computing} operations \cite{Alibart10a,pershin11d,pershin12a,linn2012beyond,diventra13a,thomas2013memristor}
 in a massively-parallel fashion and on the same physical platform where storing occurs, a striking resemblance with similar functionalities of our brain. Non-traditional computing represents an important research direction with the aim of solving the ``Von Neumann bottleneck'' that plagues
 our computer architectures \cite{Backus78a}. Moreover, memcapacitors and meminductors can store energy in the electric and magnetic field \cite{Cohen12a}, respectively, in addition to information, therefore opening new venues in the technologically important area of energy storage, distribution and manipulation \cite{dubi2011colloquium}.

 Although this whole field of research has been growing at a fast pace, there is still much confusion about the fundamental {\it physical} properties that realistic systems with memory (as opposed to ideal ones) satisfy. We believe this confusion arises from the
 different language and background of the two schools of thought that comprise the main body of researchers in this field: electrical engineers specialized in circuit theory on one side, and, on the other, physicists, materials scientists, etc. The first group often times draws conclusions from mathematical formal analogies
 without much concern about the actual practical realizations of circuit elements. The second group, instead,
 reasons with the idea that science is, first and foremost, an experimental enterprise, and as such they place the
 theoretical constructs as subordinate to experimental facts. Irrespective, it is important to stress that while mathematical analogies have their value, they do not always correspond to an objective physical reality.

 In this paper, due to our personal training as physicists, we take the second stance and refer to some fundamental physical theories that go way back in time, even before the year 1971 when the
 axiomatic definition of "memristors" was first introduced \cite{chua71a}. These theories put on a firm ground the experimental fact that {\it any} condensed matter system -- which is ultimately comprised of electrons and ions -- cannot respond instantaneously to external perturbations. This is even more so in systems of nanoscale dimensions
 where the dynamics of a few atoms may affect the whole structure dramatically~\cite{diventra08}.
 As such, some degree of memory in the {\it response} of the system to external fields is always present. This also
 shows that {\it ideal} resistors, capacitors and inductors are just circuit theory idealizations of actual
 properties of real systems, being a good representation of such properties only within a range of experimental
 conditions (e.g., within certain intervals of amplitudes and frequencies). It also shows that memristive, memcapacitive and meminductive systems \cite{diventra09a,diventra09b} are simply
 resistors, capacitors and inductors, respectively, whose memory is made more apparent under certain experimental conditions.

 Once we realize that resistances, capacitances and inductances are simply response functions, all other constraints
 introduced ``artificially'' in the mathematical (axiomatic) definition of memristors \cite{chua71a,chua76a,chua2012fourth} -- and still assumed in some of the literature dealing with memcapacitors and meminductors as well -- have no reason to exist. Such unphysical constraints are {\it i}) the finiteness of the responses themselves at all times, with consequent crossing of the input-output curve under a periodic drive, and {\it ii}) -- in the case of memcapacitors and meminductors --  positiveness of their response functions at all times.

We finally discuss some limitations of {\it ideal} memristors, memcapacitors and meminductors. By ideal memristors we mean those whose resistance depends {\it only} on the charge that flows in the system (or on the history of the voltage) \cite{chua71a,diventra09a}. Ideal memcapacitors are those whose capacitance depends only on the history of the charge stored on their plates (or the history of the voltage across them)~\cite{diventra09a}. Finally, ideal meminductors are those whose inductance only depends either on the history of the current that flows through them or their flux~\cite{diventra09a}. We show that these ideal situations do not always represent the physical reality. In particular, the state of ideal memelements does not have any energetic protection against (the unavoidable) fluctuations leading to what could be named a {\it stochastic catastrophe} of the memory state.

Such a lack of protection also violates the {\it Landauer principle} of minimal energy dissipated per logic operation~\cite{landauer61a}. In addition, ideal memelements show an {\it over-delayed switching effect} not
 observable in actual experiments, and their modeling needs to be compatible with the {\it symmetries of electrodynamics}. In other words, the ideal memelements can only be considered as approximations of actual realizations. For instance, even the much-discussed example of memristor as put forward by Hewlett-Packard a few years back \cite{strukov08a}, and which has rekindled the interest in this field, is -- in its actual
material realization -- {\it not} an ideal memristor. On the other hand, the model used in that publication {\it does} correspond
to an ideal memristor, and is thus unable to reproduce all the experimental facts.

\section{Derivation of memristive properties from Kubo response theory}

In this section we show explicitly how memristive properties emerge naturally when a system is subject to an
external field. In order to do this we rely on a well-known theory, that dates back to the 50's, due to Kubo, who derived such response -- both classical and quantum -- using perturbation theory \cite{kubo1957statistical}. Although the original publication deals also with higher-order perturbations, Kubo response theory is now mainly employed in the linear response regime. In order to keep the math at a minimum, we will also focus on the linear response case,
but the conclusions we draw are valid also in the non-linear case. In addition, it is not our goal here to re-derive the whole apparatus of Kubo's theory -- which can be found in the original publication or textbooks that expand on the subject (see, e.g., Ref. \cite{diventra08}) --, just use its salient results to make our point.

Let us then calculate the electrical {\it current density}, $\vec{j}({\bf r},t)$ of a given material when subject to an external electric field $\vec{E}({\bf r},t)$. In order to do this at the microscopic level, we assume we know the many-body electron Hamiltonian, ${\cal H}(\{\vec{R} \} )$, as a function of the atomic positions, $\{\vec{R} \} $, of all the ions in the material. For simplicity, we treat here the ionic positions classically. They thus follow a
classical Newton equation. For an ion with mass $M$ this is
\begin{equation}
\frac{\textnormal{d}^2\vec{R}_i}{\textnormal{d}t^2}=\frac{\vec{F}\left( \left\{ \vec{R} \right\}, \left\{\textnormal{d}\vec{R}/\textnormal{d}t \right\} \right)}{M}, \label{Newton}
\end{equation}
where $\vec{F}$ is the total force acting on that particular ion.

Notice that we have explicitly included in the force $\vec{F}$ also a dependence on the velocity of the ions. The reason for this is because, even at the classical level, the electron-ion and the ion-ion interactions exert a
``drag'' for an ion to move in the lattice under the action of an external field (a consequence of this is, e.g., the inelastic contribution to electromigration \cite{diventra08}).
Under these assumptions, the Hamiltonian is
then {\it parametrically} dependent on the ionic positions and velocities, which, in turn follow their own equations of motion. In
the language of memristive systems, we can anticipate that the ionic coordinates and velocities represent {\it state variables} of the system. (Clearly, in some memory systems, some electronic degrees of freedom -- such as the charges on impurity atoms -- should also be considered as internal state variables.)

With this Hamiltonian in hand, we then switch on the electric field perturbation at time $t_0$. By applying Kubo response theory to obtain the current-current response function \cite{kubo1957statistical} (whether treating the electron problem quantum-mechanically or fully classical) we then obtain ($\mu,\nu=x,y,z$)
\begin{eqnarray}
j_{\mu}({\bf r},t)= &&
\sum_{\nu}\int \textnormal{d}{\bf r}'\int_{t_0}^t\textnormal{d}t'\sigma_{\mu \nu}\left(  {\bf r}, {\bf r}';t,t'; \left\{ \vec{R} \right\},
\left\{ \frac{\textnormal{d}\vec{R}}{\textnormal{d}t} \right\} \right) \nonumber \\
&& \times
 E_{\nu}({\bf
r}',t'). \label{linear}
\end{eqnarray}
Here, the sum is over the three spatial coordinates, and $\sigma_{\mu \nu}({\bf r}, {\bf r}';t,t'; \{\vec{R} \}, \{\textnormal{d}\vec{R}/\textnormal{d}t \})$ is a 2-rank tensor representing the response (the electrical {\it conductivity} of the system) in the direction $\mu$ under an electric field component in the direction $\nu$. This response is non-local in space, as represented by the electronic coordinates ${\bf r}, {\bf r}'$, but most importantly, for the discussion that follows, it is non-local in {\it time}, namely the conductivity depends on the full {\it history} of the system from the time the
perturbation was switched on. We stress here that this memory is not just in the electronic degrees of freedom at fixed ionic positions, namely, in
the delay inherent in the effective interaction among electrons. It also
originates from the dynamics of the classical ions and consequent change of the many-electron configurations, which
in turn affect the current.

Assuming that the varying electric field induces a negligible magnetic field, we can now write $\vec{E}=-\vec{\nabla} V$, with $V$ the electric potential. The {\it total} current is $I=\int_S \vec{j}
\cdot \textnormal{d} \vec{S}$, with $ d \vec{S}$ the infinitesimal surface vector of the surface $S$ through which the current is measured. Then from Eq.~(\ref{linear}) we find (a similar derivation can be found in Ref. \cite{vignale2009incompleteness})
\begin{equation}
I(t)=G\left( \left\{ \vec{R} \right\},
\left\{ \frac{\textnormal{d}\vec{R}}{\textnormal{d}t} \right\} ,t\right)V(t), \label{GR}
\end{equation}
where the conductance $G(\{\vec{R} \},\{\textnormal{d}\vec{R}/\textnormal{d}t \},t)$ is for a two-terminal device along the $x$ direction is given by
\begin{eqnarray}
&&G\left( \left\{ \vec{R} \right\},
\left\{ \frac{\textnormal{d}\vec{R}}{\textnormal{d}t} \right\} ,t\right)= \nonumber \\
&&-\int\limits_{C_1} \textnormal{d}{\bf r}\int\limits_{C_2} \textnormal{d}{\bf r}'\int\limits_{t_0}^t\textnormal{d}t'\,\tilde{\sigma}_{xx}\left( {\bf r}, {\bf r}';t,t'; \left\{ \vec{R} \right\},\left\{ \frac{\textnormal{d}\vec{R}}{\textnormal{d}t} \right\}\right)\, \;\;\; ,
\end{eqnarray}
where the integrals are over the far-left and far-right device surfaces $C_1$ and $C_2$.

 If we now call $\{x_1 \}= \{\vec{R}  \}$ and $\{x_2 \}=\{\textnormal{d}\vec{R}/\textnormal{d}t \}$, from the Newton equation~(\ref{Newton}) the set of equations
\begin{equation}
I(t)=G( \{x_1 \}, \{x_2 \},t)V(t), \label{I_12}
\end{equation}
\begin{equation}
 \{\dot x_2  \}=\left \{ \frac{\vec{F}}{M}\right \}=f(\{x_1 \}, \{x_2 \},t),\label{x2}
\end{equation}
define a {\it memristive} system \cite{chua76a},
with $\{x_1\}$ and $\{x_2 \}$ the set of internal state variables of the system (in this particular case, the position and velocities of all ions in the material). By writing $\{ x \}= \{ x_1, x_2 \}$ the ensemble of all state variables, we would then write Eq.~(\ref{I_12}) in its most familiar form \cite{chua76a}
\begin{equation}
I(t)=G( \{x \},t)V(t) \label{I_x}.
\end{equation}

The derivation we have followed has been performed in the linear regime. A similar derivation could be carried out
by including higher orders in the perturbation expansion \cite{kubo1957statistical}. This would lead to more complicated expressions that contain explicitly the dependence of the conductivity on the electric field (and hence potential $V$) making the
conductance {\it non-linear}. Without writing these expressions explicitly (the reader can find, e.g., the second-order
expansion in the original paper \cite{kubo1957statistical}) we can definitely write the current in this case as
\begin{equation}
I(t)=G(  \{x \},V,t)V(t)\label{GRV},
\end{equation}
which together with Eqs.~(\ref{x2}) represent a memristive system \cite{chua76a}.

\section{Derivation of memcapacitive properties from Kubo response theory}

Without delving too much into the details of the calculations, one can compute the capacitance of
a given system using the above response theory approach. In this case the memory may arise directly from the
permittivity itself, as originating from the delayed response of dipoles in the dielectric of the capacitor, and/or
in the geometrical changes of the metallic plates defining the capacitor, which again can be represented as classical
equations of motion for the positions $\{\vec{R}\}$ and velocities $\{\textnormal{d}\vec{R}/\textnormal{d}t \}$ of the ions composing the metallic plates [Eqs.~(\ref{x2})]. The {\it permittivity}, $\epsilon({\bf r}, {\bf r}';t,t'; \{\vec{R}\},\{\textnormal{d}\vec{R}/\textnormal{d}t \})$ can be calculated from the {\it density-density} response function (as opposed to the
current-current response function calculation of the resistivity) \cite{kubo1957statistical}. By performing the actual calculation of the capacitance with this permittivity
would then lead to a capacitance $C( \{x \},t)$ which is state dependent. Going beyond linear response we would then obtain the final result
\begin{equation}
q(t)=C( \{x \},V,t)V(t), \label{CRV}
\end{equation}
where $q(t)$ is the total charge on the capacitor, and $V(t)$ is the potential across it. Equation~(\ref{CRV}) represents a memcapacitive system \cite{diventra09a}.

\section{Derivation of meminductive properties from microscopic theories}

The calculation of meminductive properties from a microscopic theory of magnetization is trickier because of the
complexity of the problem. Indeed, here quantum mechanics is necessary to explain magnetic phenomena in materials since a classical description cannot account for diamagnetism, paramagnetism, or even ferromagnetism \cite{ashcroft2005solid}. For this
reason, we generally proceed by using phenomenological equations for the magnetization dynamics \cite{gilbert55a},
and then derive the magnetic flux through the inductor as {\it response} to the current flowing through it. In this case, the memory of the inductance can depend on both the magnetization history as well as on the geometrical changes of the inductor \cite{pershin11a}. In fact, following these microscopic calculations of the inductance one can obtain relations of the type
\begin{equation}
\phi(t)=L( \{x \},I,t)I(t), \label{LRV}
\end{equation}
where $\phi(t)$ is the flux-linkage (integral of the voltage), $I(t)$ the current, and the inductance $L$ depends also on some
state variables with their own equations of motion.

\section{Generalized response functions}

It is now a simple matter of abstraction to generalize the above definitions by invoking
a {\it general non-linear, memory-dependent response function} $g$. Equations~(\ref{GRV}),~(\ref{CRV}), and~(\ref{LRV}), together with the equations of motion for the state variables, can then be lumped into a single type of
expressions~\cite{diventra09a}
\begin{eqnarray}
y(t)&=&g\left(x,u,t \right)u(t) \label{Geq1}\\ \dot{x}&=&f\left(
x,u,t\right) \label{Geq2}
\end{eqnarray}
with $f$ some vector function of internal state variables, and $u(t)$ and $y(t)$ the input and output signals, respectively.

At this point, due to the above derivations, we could ask the question of whether these memory elements are just a renaming of previous work, and if so what value does this have, if at all. While a case can be made regarding the renaming of certain physical features that have been studied extensively in the past -- and we should all be aware of --, the definitions of memristive, memcapacitive and meminductive systems represent an economic way of describing a huge amount of systems, materials and devices with memory in an {\it unified}, general framework. In fact, as we have also noted in other publications (see, e.g., our Ref. \cite{di2011memory}), this unified description is a source of inspiration for new ideas and concepts across different disciplines. For instance, it is worth stressing that the definition embodied in Eqs.~(\ref{Geq1}) and~(\ref{Geq2}) is not limited only to the input perturbations we have discussed so far, such as charge, current, voltage and flux. It represents
{\it any} response of a given system to an arbitrary perturbation that induces memory in the output.

Therefore, by abstracting the general definitions from the microscopic mechanisms that lead to memory, we can study more complex situations such as, for example, the collective properties of complex networks of these elements \cite{pershin13a}, without ever specifying the particular (practical) realization of such elements. Such networks with memory are not just abstract constructs. They rather represent idealized -- but nonetheless very useful -- models of complex biological behavior, including possibly some features of our brain \cite{diventra13a}.

\section{Some general properties of response functions}
Now that we have showed that memristances, memcapacitances, and meminductances are response functions with memory, we can enumerate some of the properties that they satisfy. First of all, we need to stress that
response functions are {\it not} observables: they are relations between the input and output signals, which {\it
are} observables. Therefore, from a physical point of view the input and output signals have to be bound functions
of time. This limitation does not translate to the response functions: at any given time the input $u(t)$ may be zero, while the output $y(t)$ remains finite. From Eq.~(\ref{Geq1}) it is then obvious that the response function is {\it infinite} at that particular instant. An example of this is a system which is driven into a superconducting state at
some moment of time from a metallic state: the conductance of the system goes from a finite value (metallic state) to
an infinite value (superconducting state). Therefore, unlike what has been always assumed for memristive systems
\cite{chua76a}, we do not need to artificially enforce the response function to be finite.

This property also implies another important one. If the response function can acquire an infinite value
at certain times, then it is not necessary that memristive, memcapacitive and meminductive systems -- or any other
system satisfying Eqs.~(\ref{Geq1}) and~(\ref{Geq2}) -- show ``pinched" hysteresis loops, namely at the time
 when $u$ is zero, the response $g$ may be infinite, and therefore $y$ is finite. Conversely, we may have situations in which at some time the response function is zero, the output $y$ is zero, but the input is finite. As an example, suppose that we again consider a metallic system and drive it into
a superconducting state, and then revert the dynamics to the metallic state, but following a different path. This
system has memory but its characteristics curve will not pass through the origin.

The existence of a {\it pinched} hysteresis curve has always been declared as a hallmark of memristive systems \cite{chua76a}, but
as we have just shown it is {\it neither} necessary {\it nor} physically important to characterize a system with memory.
The pinched hysteresis curve is simply a typical feature.
Similar considerations hold also for memcapacitive and meminductive systems. For example, we reported O-shaped hysteresis curves in solid-state
memcapacitive systems \cite{martinez09a}.

\begin{figure}[t]
\centering\includegraphics[width=.40\linewidth]{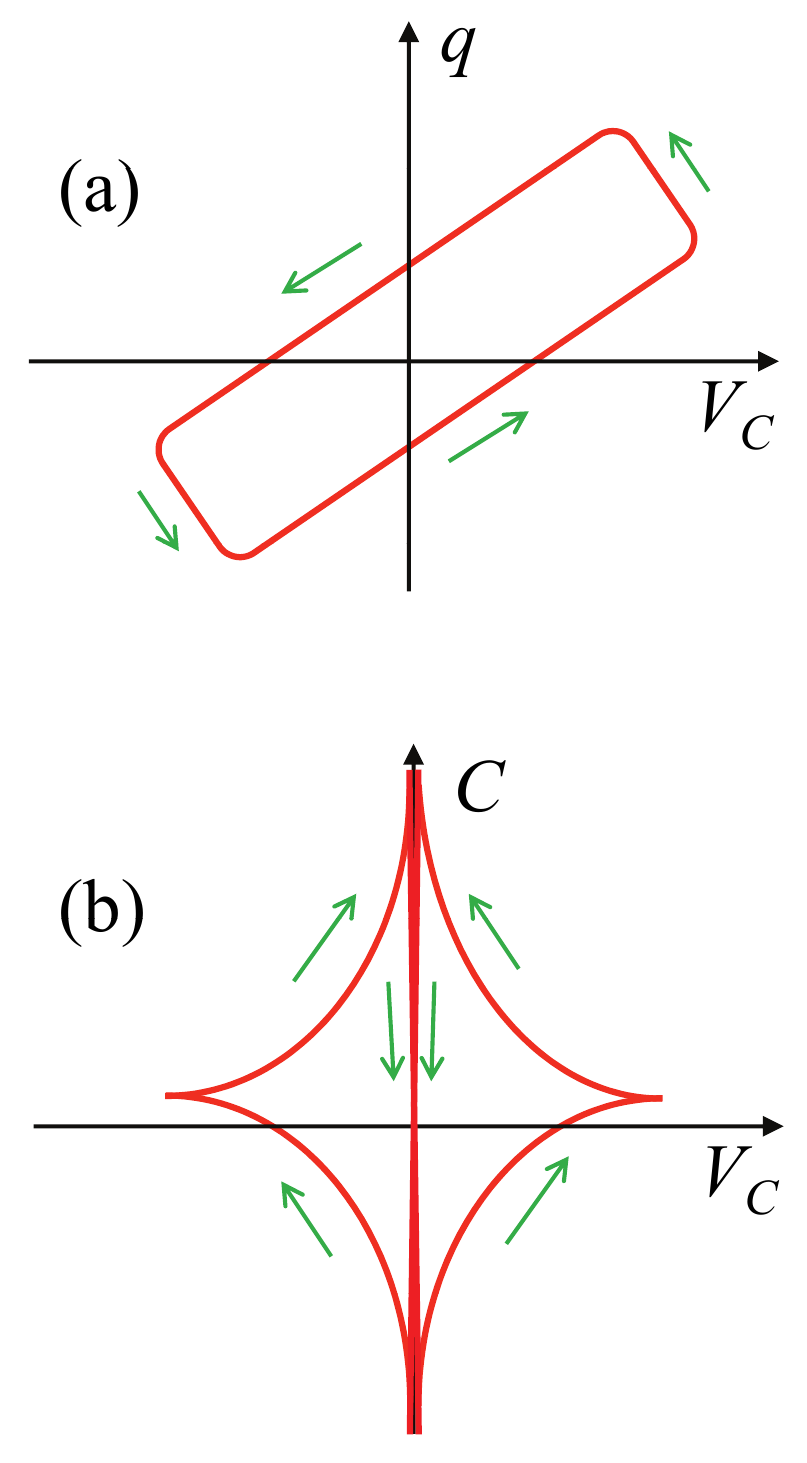}
\centering \caption{(a) Schematics of a memcapacitor charge-voltage curve that does not cross the
origin. (b) Corresponding capacitance-voltage plot showing negative and diverging capacitances.}
\label{fig1}
\end{figure}

Another important property pertains to the sign of the response function $g$ at any given time. For memristive
systems the condition of passivity implies that the resistance is {\it always} non-negative at all times (a negative
resistance can only ensue from an active element). However, such a condition does not preclude
a change of sign of the capacitance and inductance at certain times. This can be physically understood quite
easily. Take for instance a memcapacitive system, whose permittivity lags behind the voltage applied to the system. In that case, at the instants of time when the voltage changes sign, the dielectric cannot fully screen this field.
This under-screening effect results in the ``wrong'' sign of charges on the capacitor plates compared to the direction of the field, and therefore in a {\it negative capacitance}. Similar considerations can be made when the material
between the plates ``over-screens'' the field at certain instants of time. An example of negative
capacitance is given in Fig. \ref{fig1}, see also Ref. \cite{martinez09a}.

Finally, these results apply also to
meminductive systems, when, e.g., the permeability of the magnetic material cannot instantaneously follow the current (or flux)
across the inductor. Putting all this together, we can then sketch a possible hysteresis loop for memory elements when they do not cross the origin (see Fig.~\ref{fig1}).

\section{Physical limitations of ideal memelements}

Let us now focus on the physical limitations of {\it ideal} memristors, memcapacitors and meminductors. As
already mentioned, these are particular cases of the general class of memelements defined by Eqs.~(\ref{Geq1}) and~(\ref{Geq2}).

\subsection{Sensitivity to input fluctuations and stochastic catastrophe}

We first note a {\it general physical principle}: non-volatile information storage {\it without} energy barriers that separate distinct memory states is
{\it impossible}. Usually, for stability of the stored information over times much longer than any practical reading time, an energy barrier much larger than $kT$ is required \cite{landauer61a}, where $k$ is the Boltzmann constant, and $T$ is the temperature of the
environment.
The models of {\it ideal} memelements do miss such barriers. Therefore, in all these cases, even very small input signals - applied for a sufficient time - can change the system state. This implies
a high sensitivity of ideal memelements to fluctuations in the input variable. Being unprotected against
fluctuations, internal states exhibit diffusive dynamics (similar to Brownian motion \cite{gardiner2004handbook})  causing the state degradation, a phenomenon we could name {\it stochastic catastrophe}.

In order to demonstrate this feature, let us consider an ideal current-controlled memristor. Its definition implies that, in the presence of input noise, its internal state
variable $x$ is described by
\begin{equation}
\frac{\textnormal{d}x}{\textnormal{d}t}=I(t)+\xi(t) , \label{eq:memr_noise}
\end{equation}
where $\xi(t)$ is, e.g., Gaussian white noise,
\begin{equation}
\langle \xi(t)\rangle=0, \;\;\;\;\langle \xi(t)\xi(t^\prime)\rangle=2\kappa \delta(t-t^\prime).
 \label{eq23}
\end{equation}
Here, $\kappa$ is a positive constant characterizing the noise strength. Integrating both sides of Eq. (\ref{eq:memr_noise}) gives
\begin{equation}
x(t)-x(0)=q(t)+\int\limits_0^t\xi(t^\prime)\textnormal{d}t^\prime. \label{eq24}
\end{equation}
Moving $q(t)$ to the left-hand side of Eq. (\ref{eq24}), squaring both sides, and averaging over
the ensemble realizations of the noise, we find
\begin{equation}
\langle (x(t)-x(0)-q(t))^2 \rangle =\int\limits_0^t\int\limits_0^t\ \langle  \xi(t^\prime) \xi(t^{\prime\prime}) \rangle \textnormal{d}t^\prime \textnormal{d}t^{\prime\prime}=2 \kappa t, \label{eq25}
\end{equation}
which shows that the characteristic deviation of the internal state variable from the deterministic trajectory $x(t)=x(0)+q(t)$ increases as the square root of time like for a Brownian particle \cite{gardiner2004handbook}. Similar considerations apply to all types of {\it ideal} memelements.

In particular, the (intrinsic) thermal agitation of electrons inside any
voltage-controlled memristors is responsible for the well-known Johnson-Nyquist noise \cite{johnson1928thermal,nyquist1928thermal} (voltage fluctuations). These fluctuations are present regardless of any applied voltage, and even in systems that are not connected to any circuit at all. The thermal voltage fluctuations thus act as an internal degradation mechanism in such devices, which in the
absence of any energy barrier to protect the state of the system, leads to a {\it diffusive loss of information}.

\subsection{Violation of Landauer principle}

Moreover, the logical (and hence physical) irreversibility of any computing machine imposes a minimal heat generation condition on any memory device \cite{landauer61a}. This minimal heat generation is of order $kT$ per machine
cycle, and is known as {\it Landauer principle}~\cite{landauer61a}.
The satisfaction of this condition in memristors has been indeed questioned in a recent publication \cite{meuffels2012fundamental}.

To understand the reasoning behind Ref. \cite{meuffels2012fundamental} conclusion, let us consider the switching of a memristor
at constant temperature and pressure. Under these conditions, the relevant thermodynamic potential is the Gibbs free energy, which should involve
energy barriers between different information states and corresponding heat dissipation as suggested by Landauer \cite{landauer61a}. However, the equation of memristor dynamics, such as Eq. (\ref{eq:memr_noise}), does not involve any restrictions on minimal switching energy, and thus violates Landauer's principle. In fact, this is simply another consequence of not having energy barriers between different memory states in any ideal memristor model. Note that
although Landauer's principle was formulated for digital computing, the same physical constraints apply also to
analog computing as well, such as the one that can be performed with memory elements~\cite{pershin12a,diventra13a}.

\subsection{Over-delayed switching}

Another shortcoming of {\it ideal} memelements is related to the {\it over-delayed switching effect} not observable in realistic memdevices. Consider, for example, an ideal current-controlled memristor described by the resistance relation $R=R(q)$. If the switching of this memristor occurs in the vicinity of $q=0$ and the device is subsequently placed in a state with a large $q$, then a charge $-q$ should flow
through the device for it to switch back. From experiments, however, we know that the switching actually occurs as soon as the applied voltage or current exceeds its threshold value \cite{waser07a,Jo09a,pershin11a}.
Thus, in real devices, a much smaller amount of charge $-q'$ would be enough to switch the memristor back ($q'\ll q$). In other words, real devices do not really "track" the charge flown through them when they are in their limiting states (e.g., ON and OFF memristance states). Indeed, this limitation of ideal models should be considered in time-dependent simulations of realistic systems.

\subsection{Incompatibility with symmetries of electrodynamics}

Finally, we would like to emphasize that the symmetry used to postulate the memristor \cite{chua71a} is {\it not} a symmetry of electrodynamics. Electrodynamics is governed by Maxwell's equations that are invariant under {\it charge conjugation} ($q\rightarrow -q$), {\it parity} ($\vec{r}\rightarrow -\vec{r}$) and {\it time reversal} ($t\rightarrow -t$) transformations \cite{bigi1999cp}.
Any electronic device and its models (if the device operation is based solely on microscopic electrodynamics) should satisfy these symmetries. Note, however, that resistors (as well as memristors, memristive systems, and any dissipative memelements), violate time-reversal invariance: The operation of such elements involves dissipation -- the conversion of electric potential energy into heat -- that can not be reversed by changing the arrow of time.

On the
other hand, the other symmetries of electrodynamics - in particular, charge conjugation and parity - do need
to be satisfied. For instance, for an ideal memristor, charge conjugation requires $R(q)=R(-q)$, in
such a way that Eq.~(\ref{GRV}) (which in this case is simply $V(t)=R(q)I(t)$) be invariant, namely
the resistance needs to be an even function of the charge. This is definitely not taken into account
in many simple models of ideal memristors~\cite{strukov08a}.

Note also that for the more realistic cases of memristive, memcapacitive, and meminductive elements, the relevant symmetries of electrodynamics
are generally satisfied by the physical requirements that lead to the equations of motion of the internal state
variables~\cite{pershin11a}.

\section{Conclusions}

In conclusion, by using the modern theory of response functions (both in the quantum and classical regimes) as developed in the 1950s by Kubo \cite{kubo1957statistical}, we have shown that memristances, memcapacitances and meminductances (that describe different devices with memory) are simply response
functions. As such they have to satisfy well-defined physical properties. Consequently, any additional artificial limitation, such as finiteness of these responses -- introduced, for instance, in the axiomatic definition of memristive systems \cite{chua71a,chua76a,chua2012fourth} -- limit the range of possible physical responses.

We have also
discussed several potential problems related to the axiomatic definition of {\it ideal} memory devices, such as
the stochastic catastrophe, violation of Landauer principle, over-delayed switching, and possible
incompatibility with symmetries of electrodynamics. These limitations should be taken into account in circuit simulations. In fact, more realistic device models -
corresponding to the more general class of memristive, memcapacitive and meminductive systems - that include the actual physics of the device operation should be used in practice. We hope this paper clarifies many misunderstandings that
are still propagated in the literature.

\ack

This work has been partially supported by NSF grants No. DMR-0802830 and ECCS-1202383, and the Center for Magnetic Recording Research at UCSD.

\section*{References}
\bibliographystyle{unsrt}
\bibliography{memcapacitor}
\end{document}